\documentclass[10pt,conference,a4paper]{IEEEtran}
\IEEEoverridecommandlockouts
\usepackage{cite}
\usepackage{amsmath,amssymb,amsfonts}
\usepackage{algorithmic}
\usepackage{graphicx}
\usepackage{textcomp}
\usepackage{xcolor}
\usepackage{bm}
\usepackage{pgfplots}
\usepackage{flushend}
\usepackage{hyperref}
\usepackage[ruled,vlined]{algorithm2e}
\usepackage{tikz}
\DeclareUnicodeCharacter{2212}{−}
\usepgfplotslibrary{groupplots,dateplot}
\usetikzlibrary{patterns,shapes,arrows,arrows.meta}
\pgfplotsset{compat=newest}
\usetikzlibrary{positioning}

\graphicspath{{figures/}}

\DeclareMathOperator*{\argmax}{\arg\!\max}

\newcommand{\fig}{Fig.\ }
\newcommand{\tab}{Table }

\newcommand{\AdjacencyMatrix}{\bm{A}}
\newcommand{\AdjacencyMatrixElement}{a}
\newcommand{\FilterValues}{\bm{F}}
\newcommand{\FilterValuesElement}{f}
\newcommand{\FilterValuesVector}{\bm{\FilterValuesElement}}
\newcommand{\FilterCurve}{c}
\newcommand{\NumIterations}{I}
\newcommand{\NumChannels}{K}
\newcommand{\NumSpectra}{M}
\newcommand{\NumFilters}{N}
\newcommand{\SelectionElement}{s}
\newcommand{\Selection}{\bm{\SelectionElement}}
\newcommand{\Spectrum}{l}
\newcommand{\SpectralAngle}{\theta}

\def\BibTeX{{\rm B\kern-.05em{\sc i\kern-.025em b}\kern-.08em
    T\kern-.1667em\lower.7ex\hbox{E}\kern-.125emX}}

\makeatletter
\def\ps@IEEEtitlepagestyle{%
	\def\@oddfoot{\mycopyrightnotice}%
	\def\@oddhead{\hbox{}\@IEEEheaderstyle\leftmark\hfil\thepage}\relax
	\def\@evenhead{\@IEEEheaderstyle\thepage\hfil\leftmark\hbox{}}\relax
	\def\@evenfoot{}%
}
\def\mycopyrightnotice{%
	\begin{minipage}{\textwidth}
		\scriptsize
		\copyright 2022 IEEE.  Personal use of this material is permitted. Permission from
		IEEE must be obtained for all other uses, in any current or future
		media, including reprinting/republishing this material for advertising
		or promotional purposes, creating new collective works, for resale or
		redistribution to servers or lists, or reuse of any copyrighted
		component of this work in other works. DOI: \url{https://doi.org/10.1109/MMSP55362.2022.9949059} (MMSP 2022)
	\end{minipage}
}
\makeatother
\begin{document}

\title{Optimal Filter Selection for Multispectral Object Classification Using Fast Binary Search\\
\thanks{The authors gratefully acknowledge that this work has been supported by
the Deutsche Forschungsgemeinschaft (DFG, German Research Foundation) under project number 491814627.}
}

\author{\IEEEauthorblockN{Frank Sippel, Jürgen Seiler, and André Kaup}
\IEEEauthorblockA{\textit{Multimedia Communications and Signal Processing} \\
\textit{Friedrich-Alexander University Erlangen-Nürnberg (FAU)}\\
Cauerstr. 7, 91058 Erlangen, Germany \\
\{frank.sippel, juergen.seiler, andre.kaup\} @fau.de}
}

\maketitle

\begin{abstract}
    When designing multispectral imaging systems for classifying different spectra it is necessary to choose a small number of filters from a set with several hundred different ones.
    Tackling this problem by full search leads to a tremendous number of possibilities to check and is NP-hard.
    In this paper we introduce a novel fast binary search for optimal filter selection that guarantees a minimum distance metric between the different spectra to classify.
    In our experiments, this procedure reaches the same optimal solution as with full search at much lower complexity.
    The desired number of filters influences the full search in factorial order while the fast binary search stays constant.
	Thus, fast binary search allows to find the optimal solution of all combinations in an adequate amount of time and avoids prevailing heuristics.
    Moreover, our fast binary search algorithm outperforms other filter selection techniques in terms of misclassified spectra in a real-world classification problem.
\end{abstract}


\section{Introduction}

\label{sec:intro}

Multispectral imaging (MSI) systems are getting more and more popular in recent years.
Especially for classification MSI is often employed, since many different materials absorb light at characteristic wavelengths.
For example, MSI systems can be used in food processing to determine the amount of sugar in melons \cite{sugiyama_visualization_2010}.
Moreover, these cameras can be used in environmental engineering for classifying different types of plastics \cite{moroni_pet_2015}, and in medical environments, for example, when estimating the degree of burn injuries \cite{sowa_classification_2006}.

\begin{figure}[t]
	\begin{center}
		\includegraphics[scale=0.8]{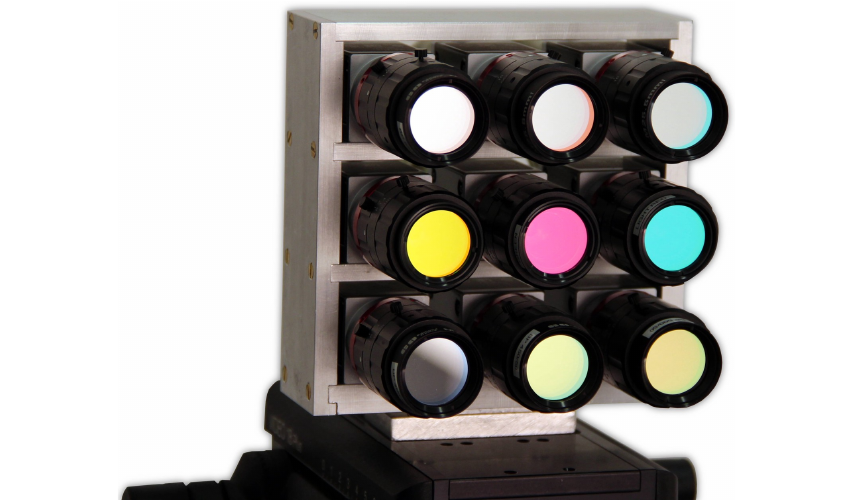}
	\end{center}
	\caption{A multispectral camera with nine channels \cite{genser_camera_2020}.}
	\label{fig:ms_camera}
\end{figure}

Different setups for multispectral cameras are possible.
One approach is by using a multi-camera setup \cite{genser_camera_2020}, which is shown in \fig\ref{fig:ms_camera}.
Here, nine cameras are employed, where each of the cameras has a filter mounted in front of the camera lens.
Another possibility to build an MSI camera is by using a filter wheel.
Here, only one camera is needed and a wheel with filters attached to it is mounted in front of the camera \cite{brauers_multispectral_2008}.
When imaging the object, the wheel is rotated such that each filter is in front of the camera once.
MSI images cannot only be recorded by putting the filters in front of the camera but also by filtering the imaging light when recording in a controllable environment.
One way is by placing similar filters in front of the light \cite{chi_multi-spectral_2010}.
Another possibility is multiplexed illumination, where several filtered LEDs are mounted to a panel and a subset of the different types are turned on, when recording one image \cite{park_multispectral_2007}.
After several of these images are recorded, the light spectrum for each pixel is reconstructed in the end.
This can be done, for example, by using guided filtering \cite{sippel_structure-preserving_2020}.

In comparison to hyperspectral imaging (HSI), which aims at sampling the light spectrum equidistantly for each pixel, MSI has less channels and is therefore much cheaper.
However, while the filters of an HSI system are implicitly fixed, the filters of an MSI can be freely chosen from a set of available filters.
The differentiating properties of these filters include bandwidth, shape, and central wavelength.
Thus, an important question is how to choose these filters optimally for the desired application from a set of available filters.
Unfortunately, a full search (FS) quickly gets infeasible.
For example, when selecting nine filters for the camera in \fig\ref{fig:ms_camera} from 150 available filters, the number of possibilities to combine them is $\left(\begin{array}{c}150 \\ 9 \end{array}\right) = 82947113349100$.
Evaluating nearly 83 trillion different filter possibilities is practically infeasible.
Thus, a method to approximate the best set of filters for the underlying application in an adequate amount of time is highly desirable.

\begin{figure*}[t]
\input{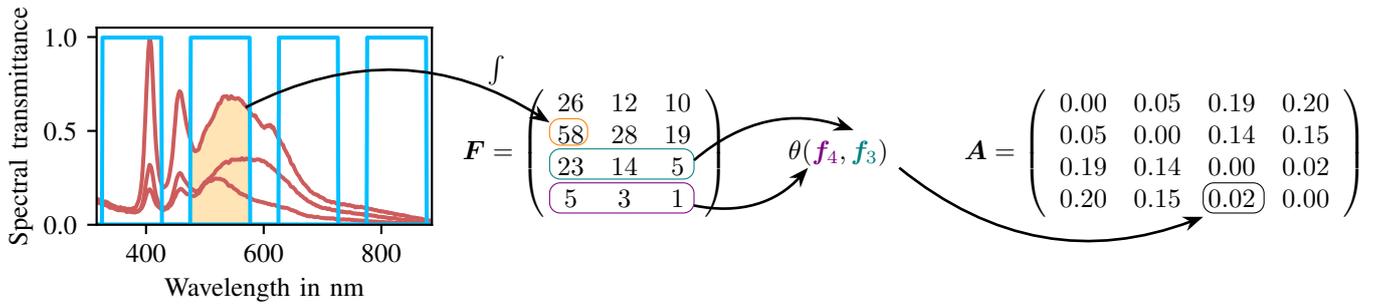}
\begin{tikzpicture}[overlay]
	\path[-{Stealth[scale=1]}, line width=1pt] (-2.6, 2.6) edge[bend left] (1.4, 2.4); \node[] at (0.7, 3.1) {$\int$};

	\node[] at (2, 2) {$\FilterValues = \left(
		\begin{array}{ccc}
			26 & 12 & 10 \\
			58 & 28 & 19 \\
			23 & 14 & 5 \\
			5 & 3 & 1
		\end{array}\right)$};
	\draw[rounded corners, color=orange] (1.4, 2.1) rectangle (1.9, 2.45) {};

	\draw[rounded corners, color=violet] (1.4, 1.2) rectangle (3.3, 1.6) {};
	\draw[rounded corners, color=teal] (1.4, 1.65) rectangle (3.3, 2.05) {};

	\path[-{Stealth[scale=1]}, line width=1pt] (3.3, 1.9) edge[bend left] (5.4, 2.3);
	\path[-{Stealth[scale=1]}, line width=1pt] (3.3, 1.3) edge[bend right] (4.8, 1.8);

	\node[] at (5.2, 2) {$\SpectralAngle(\textcolor{violet}{\FilterValuesVector_4}, \textcolor{teal}{\FilterValuesVector_3})$};

	\path[-{Stealth[scale=1]}, line width=1pt] (6, 1.8) edge[bend right] (10, 1.15);

	\node[] at (9.5, 2) {$\AdjacencyMatrix = \left(
		\begin{array}{cccc}
			0.00 & 0.05 & 0.19 & 0.20 \\
			0.05 & 0.00 & 0.14 & 0.15 \\
			0.19 & 0.14 & 0.00 & 0.02 \\
			0.20 & 0.15 & 0.02 & 0.00
		\end{array}\right)$};

	\draw[rounded corners] (10, 1.2) rectangle (10.8, 1.6) {};
\end{tikzpicture}
\caption{The pipeline from spectra and filter curves to matrix $\FilterValues$ and adjacency matrix $\AdjacencyMatrix$.}
\label{fig:rec_proc}
\end{figure*}

Several techniques exist in literature to optimize the selection of the filters.
In \cite{novati_selection_2003} a statistical analysis on recorded representative targets using all available filters is performed.
In \cite{hardeberg_filter} the orthogonality between different filters using a recorded set of the most significant reflectances is maximized.
Wang et al. \cite{wang_multispectral} examine whether narrow or wide band filters are more suitable for an MSI system.
A general problem with these techniques is that they try to optimize the filters for a general purpose camera by evaluating the resulting filter set using spectral reconstruction, which weights each wavelength equally during reconstruction.
This is not a useful procedure for selecting filters for a single application, where certain areas of the spectrum are much more useful than others for discriminating spectra.
Moreover, often whole hyperspectral images are assumed to be known for classification \cite{yang_fs-net_2021, fu_joint_2020}, which is a different problem setup since we only consider a single reference spectrum for each object without spatial information.
Another common method to choose the filters for a desired application is to handpick filters belonging to wavelength regions which appear to be heuristically interesting \cite{axelsson_detection_2013}.
Of course, this relies on the experience of the person that handpicks the filters.
Thus, to avoid this unknown, it is useful to have a system that does not depend on any experience or heuristics.
Certainly, this system has to output a filter configuration in a comparable amount of time.
Furthermore, it is also possible to just cover the whole spectral area of interest with wide bandpass filters \cite{egiazarian_multispectral_2014}.

This paper introduces a novel method to find a near optimum solution in an adequate amount of time.
Thus, no exhaustive FS has to be applied.
To the best of our knowledge, this paper is the first in literature that introduces an approximation of the full search of the filter set in the context of multispectral object classification.

\section{Proposed Method}
\label{sec:method}

The basic setup consists of $\NumSpectra$ light spectra of objects to discriminate.
For this our method Fast Binary Search (FBS) shall select $\NumChannels$ filters from $\NumFilters$ available filters.
Thus, since a camera integrates over the light spectrum that falls onto the sensor by counting photons, only the integrated values of these $\NumFilters$ filters need to be considered.
The spectral behavior of the lens and the camera are assumed to be perfect.
Otherwise, these spectral transmittance curves should be multiplied to the light spectra of the objects.
The value at row $i$ and column $j$ in the filter matrix $\FilterValues$ is calculated by
\begin{equation}
	\label{eq:integrate}
	\FilterValuesElement_{ij} = \int_{\lambda_\text{min}}^{\lambda_\text{max}} \FilterCurve_i(\lambda) \cdot \Spectrum_j(\lambda) d\lambda,
\end{equation}
where $\FilterCurve_i(\lambda)$ is the filter curve of the $i$-th available filter, $\Spectrum_j(\lambda)$ is the light spectrum of the $j$-th object to classify, and $\lambda_\text{min}$ and $\lambda_\text{max}$ are the integration bounds.
Since filters exhibit a bandpass characteristic, the bounds are typically determined by the filters.

The next step is to build an undirected graph that shows the distance between the different filters.
As distance metric, the spectral angle \cite{kruse_spectral_1993}
\begin{equation}
	\label{eq:adjacency}
	\SpectralAngle(\FilterValuesVector_{i}, \FilterValuesVector_{j}) = \arccos{\left(\frac{\FilterValuesVector_{i}^{\text{T}}}{||\FilterValuesVector_{i}||_2}\frac{\FilterValuesVector_{j}}{||\FilterValuesVector_{j}||_2}\right)}
\end{equation}
is used, where $\FilterValuesVector_{i}$ is the $i$-th row vector of value matrix $\FilterValues$.
This vector consists of all integrated values for each object for a single filter.
Thus, if two filters contain nearly the same values for all objects to classify, the spectral angle is nearly zero.
Therefore, these two filters should rather not be used together.
With this spectral angle, the adjacency matrix $\AdjacencyMatrix$ of the undirected graph can be built by evaluating the spectral angle for each filter pair combination $\AdjacencyMatrixElement_{ij} = \SpectralAngle(\FilterValuesVector_{i}, \FilterValuesVector_{j})$.
This process is summarized in \fig\ref{fig:rec_proc}.
Note that any distance metric can be used in this framework, e.g. the spectral information divergence \cite{chang_spectral_1999} or the spectral correlation angle \cite{carvalho_approach_2011}.

The goal is to maximize the minimum spectral angle of a configuration with $\NumChannels$ filters.
Thus we want to find the $\NumChannels$ nodes in the undirected graph which maximize the minimal distance within the graph.
For this, a binary search over the spectral angle is employed, since the maximum number of connected nodes with the current minimal spectral angle is monotonically decreasing with an increasing minimal spectral angle.
The starting bounds for the binary search are $\SpectralAngle_{\text{min}} = \min(\AdjacencyMatrix)$ and $\SpectralAngle_{\text{max}} = \max(\AdjacencyMatrix)$, respectively.
Consequently, within the binary search procedure, the current tested minimal spectral angle is
\begin{equation}
	\SpectralAngle_{\text{curr}} = \frac{\theta_{\text{min}} + \SpectralAngle_{\text{max}}}{2}.
\end{equation}
The next step for FBS is to find the maximum number of connected nodes possible with the current minimal spectral angle to know whether a configuration for the current spectral angle exists for which enough filters can be picked.
For this, a mixed integer maximization problem \cite{wah_mixed_2008} is set up.
The objective is to maximize the number of nodes that can be chosen for the current minimal spectral angle.
As soon as edge weights fall under the current minimal spectral angle, a constraint is added to the optimization problem that prevents picking the filters that are connected by this edge at the same time.
Thus, the mixed integer maximization problem can be written as
\begin{equation}
	\begin{aligned}
		\hat{\Selection} = &\argmax_{\Selection} \ \sum_{i=1}^{\NumFilters} \SelectionElement_i\\
		& \ \text{s.t.} \quad \SelectionElement_i + \SelectionElement_j <= 1 \quad \forall i \neq j,  \AdjacencyMatrixElement_{ij} < \SpectralAngle_{\text{curr}},
	\end{aligned}
\end{equation}
where $\Selection$ is the binary selection vector.
Thus, only values 0 or 1 are allowed.
Afterwards, depending on the resulting value the new spectral angle bounds for the next iteration can be set.
If the resulting sum is lower than the desired number of filters, the upper bound of the binary search is set to the current minimal spectral angle, since it is known that no valid configuration can be found above the current spectral angle.
On the other hand, the lower bound is set to the current minimal spectral angle if the resulting sum is greater or equal to the number of desired filters
\begin{equation}
	\begin{cases}
		\SpectralAngle_{\text{min}} = \SpectralAngle_{\text{curr}} & \text{if} \ \sum_{i=1}^{\NumFilters} \SelectionElement_i >= \NumChannels \\
		\SpectralAngle_{\text{max}} = \SpectralAngle_{\text{curr}} & \text{else}.
	\end{cases}
\end{equation}
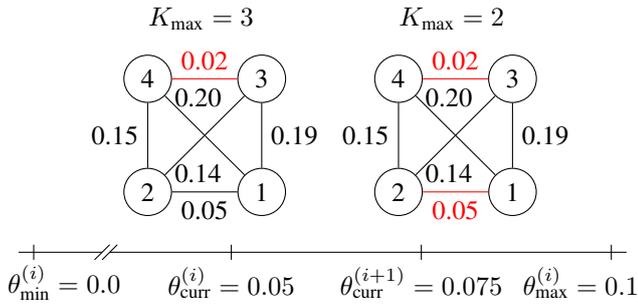
\begin{figure}[t]
\begin{tikzpicture}
	\draw (0, 0) -- (1.1, 0);
	\draw (1.2, 0) -- (8, 0);

	\draw (1, -0.1) -- (1.2, 0.1);
	\draw (1.1, -0.1) -- (1.3, 0.1);

	\draw (0.2, -0.1) -- (0.2, 0.1);
	\node[] at (0.6, -0.4) {$\SpectralAngle_\text{min}^{(i)} = 0.0$};
	\draw (7.8, -0.1) -- (7.8, 0.1);
	\node[] at (7.4, -0.4) {$\SpectralAngle_\text{max}^{(i)} = 0.1$};
	\draw (2.8, -0.1) -- (2.8, 0.1);
	\node[] at (2.8, -0.4) {$\SpectralAngle_\text{curr}^{(i)} = 0.05$};
	\draw (5.3, -0.1) -- (5.3, 0.1);
	\node[] at (5.3, -0.4) {$\SpectralAngle_\text{curr}^{(i + 1)} = 0.075$};

	\begin{scope}[shift={(1.7, 0.8)}]
		\node[draw, circle] (A) at (1.5, 0) {1};
		\node[draw, circle] (B) at (0, 0) {2};
		\node[draw, circle] (C) at (1.5, 1.5) {3};
		\node[draw, circle] (D) at (0, 1.5) {4};

		\draw (A) -- node[below] {0.05} ++(B);
		\draw (A) -- node[right] {0.19} ++(C);
		\draw (A) -- node[near end, above, xshift=5pt] {0.20} ++(D);
		\draw (B) -- node[near start, below, xshift=5pt] {0.14} ++(C);
		\draw (B) -- node[left] {0.15} ++(D);
		\draw[color=red] (C) -- node[above] {0.02} ++(D);

		\node[] at (0.7, 2.3) {$\NumChannels_\text{max} = 3$};
	\end{scope}
	\begin{scope}[shift={(5, 0.8)}]
		\node[draw, circle] (A) at (1.5, 0) {1};
		\node[draw, circle] (B) at (0, 0) {2};
		\node[draw, circle] (C) at (1.5, 1.5) {3};
		\node[draw, circle] (D) at (0, 1.5) {4};

		\draw[color=red] (A) -- node[below] {0.05} ++(B);
		\draw (A) -- node[right] {0.19} ++(C);
		\draw (A) -- node[near end, above, xshift=5pt] {0.20} ++(D);
		\draw (B) -- node[near start, below, xshift=5pt] {0.14} ++(C);
		\draw (B) -- node[left] {0.15} ++(D);
		\draw[color=red] (C) -- node[above] {0.02} ++(D);

		\node[] at (0.7, 2.3) {$\NumChannels_\text{max} = 2$};
	\end{scope}

\end{tikzpicture}
\caption{Two iterations of the binary search for best configuration with 3 slots.}
\label{fig:fbs_step}
\end{figure}
An example step for this binary search is given in \fig\ref{fig:fbs_step}.
Here, the three best filters out of four are searched.
The horizontal axis depicts the minimal spectral angle which is tried to be maximized.
Since we have four filters, the graphs have four nodes.
Each pair of filters is connected with the weight being the similarity of two filters regarding the spectra to classify in terms of spectral angle, see \eqref{eq:adjacency}.
If a weight falls below the current angle, this connection cannot be used and the filters are not allowed to be picked simultaneously.
Afterwards, the maximum number of connectable nodes is calculated.
For iteration $i$, the selected nodes could be $1, 2, 4$.
Since three filters are searched, the desired minimum spectral angle is increased to $0.075$ for the iteration $i+1$ according to the binary search.
FBS is summarized in Algorithm \ref{alg:algo}.

\begin{algorithm}[t]
	\SetAlgoLined
	\KwResult{Filter selection $\Selection$}

	Integrate over each spectrum and filter (see \eqref{eq:integrate})\;
	Build adjacency matrix $\AdjacencyMatrix$ (see \eqref{eq:adjacency})\;
	$\SpectralAngle_{\text{min}} = \min(\AdjacencyMatrix)$\;
	$\SpectralAngle_{\text{max}} = \max(\AdjacencyMatrix)$\;

	\While{$k < \NumIterations$}{
		$\SpectralAngle_{\text{curr}} = \left(\theta_{\text{min}} + \SpectralAngle_{\text{max}}\right)/2$

		\For{$i < \NumFilters$}{
			\For{$j < \NumFilters$}{
				\If{$\AdjacencyMatrix_{ij} < \SpectralAngle_{\text{curr}}$}{
					Add constraint $\Selection_i + \Selection_j <= 1$\;
				}
			}
		}
		Objective: $\max \ \textbf{1}^{\text{T}} \Selection$\;
		Solve MIP problem: MIP(objective, constraints)\;
		\eIf{$\textbf{1}^{\text{T}} \Selection >= \NumChannels$}{
			$\SpectralAngle_{\text{min}} =\SpectralAngle_{\text{curr}}$
		}{
			$\SpectralAngle_{\text{max}} =\SpectralAngle_{\text{curr}}$
		}

		$k = k + 1$\;
	}

	\caption{The basic procedure of FBS.}
	\label{alg:algo}
\end{algorithm}

\section{Verification}
\label{sec:experiment}

To verify the quality of the selection made by FBS, the experiment is divided into multiple sections.
First, our method is compared to a FS approach in terms of quality of the result and runtime.
Afterwards a classification experiment is conducted.

\begin{figure}[t]
	\input{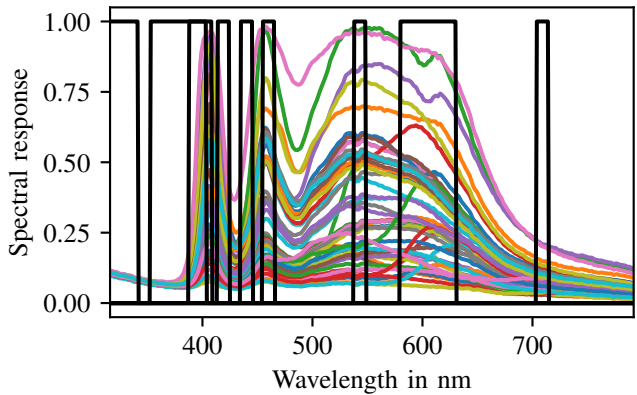}

	\caption{All spectra of different objects that shall be separated visualized by different colors and the result of FBS in black.}
	\label{fig:exp_result}
\end{figure}

For both sections, an evaluation dataset is required.
This paper uses the dataset SMM of \cite{erickson_classification_2019}, which contains 100 spectra for each of the 50 different objects.
Thus, this results in 5000 different spectra.
Since the last 95 nm in the infrared are dominated by noise and do not contain any valuable information, this spectral area is removed.
The first 25 spectra in the dataset for each object is taken for the filter optimization procedure.
These spectra are averaged to lower the influence of noise and are depicted in \fig\ref{fig:exp_result}.
The set of filters used for the quality and runtime evaluation consists of 40 bandpass filters including two different bandwidths (10 nm and 50 nm) in the range from 316 nm to 791 nm.
The number of iterations is set to 20.
Thus, the original range between the minimum and maximum spectral angle in the binary search is decreased by the factor $0.5^{20} = 9.54 \cdot 10^{-7}$.

To confirm that FBS comes close to the FS solution, different full search feasible amounts of filters are selected by FS and FBS. Afterwards the minimal spectral angle between these solution is compared as well as the selection.

\begin{figure}[t]
	\begin{center}
		\input{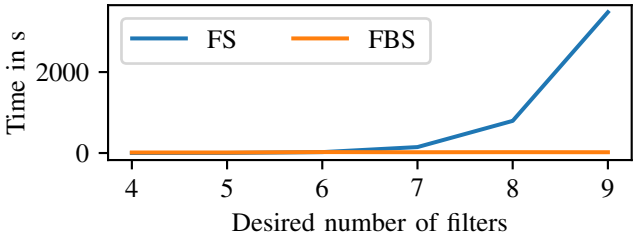}
	\end{center}
	\caption{Different runtimes for different numbers of filters.} 
	\label{fig:runtime}
\end{figure}

\tab\ref{tab:eval_diff} shows the minimal spectral angle and the selection results for the novel FBS solution and the FS search.
Since the number of filters to select from and to choose is relatively small, the FS solution can be computed.
One can see that the results in terms of minimal spectral angle are exactly the same.
Just the chosen filters differ slightly.
This results from the fact that both solutions for each evaluated number of filters are equivalent in terms of minimal spectral angle.
Therefore, this difference mainly results from the picking order of the mixed-integer program and the FS search.
The decreasing minimal spectral angle for a larger number of selected filters is reasonable.
Imagine picking two, three, and four optimized points in terms of minimal angle in a 2D-plane.
The angle will also decrease.
Figure \ref{fig:runtime} shows the runtimes of the FBS solution, which turns out to be optimal in all tested cases, and of the FS solution.
Due to the overhead of setting up the optimization problem, the runtime for a small number of selected filters is worse than of the FS solution.
However, with an increasing number of selected filters, the nature of a FS solution in such a setup is revealed.
The runtime of the FBS solution for nine desired filters is 19.6 seconds in comparison to 3479 seconds of the FS solution, which is a factor of approximately 178.
Moreover, the runtime of the mixed-integer program stays nearly constant throughout the experiment.
\begin{table}
	\centering
	\renewcommand{\arraystretch}{1.5}
	\caption{The spectral angles and the selection made by FBS and the FS solution for different numbers $\NumChannels$ of filters to select.}
	\label{tab:eval_diff}
	\begin{tabular}{c|cccccc}
		$\NumChannels$ & 4 & 5 & 6 & 7 & 8 & 9 \\
		\hline
		FS & 0.305 & 0.265 & 0.225 & 0.202 & 0.150 & 0.132 \\
		FBS & 0.305 & 0.265 & 0.225 & 0.202 & 0.150 & 0.132 \\
	\end{tabular}
\end{table}

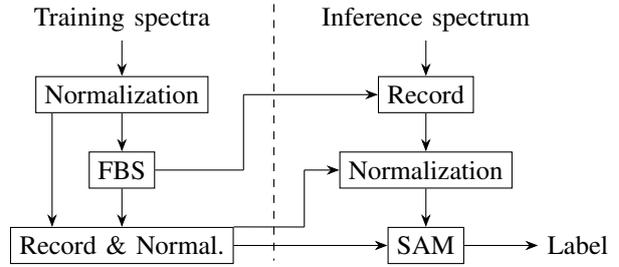
\begin{figure}[t]
	\begin{tikzpicture}[y=-1cm]

		\node[] (ts) at (0,0) {Training spectra};
		\node[draw] (n1) at (0,1) {Normalization};
		\node[draw] (fbs) at (0,2) {FBS};
		\node[draw] (n2) at (0,3) {Record \& Normal.};
		\node[] (vs) at (4,0) {Inference spectrum};
		\node[draw] (fi) at (4,1) {Record};
		\node[draw] (n3) at (4,2) {Normalization};
		\node[draw] (sam) at (4,3) {SAM};
		\node[] (lab) at (6,3) {Label};

		\draw[dashed] (2, -0.2) -- (2, 3.2);
		\draw[-{Stealth[scale=1]}] (ts.south) -- (n1.north);
		\draw[-{Stealth[scale=1]}] (n1.south) -- (fbs.north);
		\draw[-{Stealth[scale=1]}] (fbs.south) -- (n2.north);
		\draw[-{Stealth[scale=1]}] (n1.195) -- (n2.165);

		\draw[] (fbs.east) -- (1.6,2);
		\draw[] (1.6,2) -- (1.6,1);
		\draw[-{Stealth[scale=1]}] (1.6,1) -- (fi.west);

		\draw[] (n2.north east) -- (2.4,2.755);
		\draw[] (2.4,2.755) -- (2.4,2);
		\draw[-{Stealth[scale=1]}] (2.4,2) -- (n3.west);
		\draw[-{Stealth[scale=1]}] (n2.east) -- (sam.west);

		\draw[-{Stealth[scale=1]}] (vs.south) -- (fi.north);
		\draw[-{Stealth[scale=1]}] (fi.south) -- (n3.north);
		\draw[-{Stealth[scale=1]}] (n3.south) -- (sam.north);
		\draw[-{Stealth[scale=1]}] (sam.east) -- (lab.west);
	\end{tikzpicture}
	\caption{The filter selection and classification procedure of FBS for the second experiment.}
	\label{fig:bd_classification}
\end{figure}

To test the actual capability, a setup is chosen where 9 filters out of 72 different filters are selected, which are uniformly distributed in the same wavelength range.
This time, the bandwidths of the different filters are 10 nm, 20 nm and 50 nm.
The block diagram for the training and inference procedure is shown in \fig\ref{fig:bd_classification}.
First, the training spectra are normalized such that a scale factor on the spectrum, which might occur due to different lighting and reflectance conditions, does not influence the FBS procedure.
Afterwards, FBS is executed to find suitable filters.
The result of the optimization procedure is shown in \fig\ref{fig:exp_result}.
The chosen filters are nicely distributed across the whole wavelength range and interesting areas to classify the different spectra are covered.
The result may be not very intuitive, since a human tends to place the filters at locations where the order of spectra changes.
However, the maximum spectral angle is achieved in regions where the order changed maximally, which are usually rather flat.
Furthermore, the spectral angle used for FBS normalizes the recorded value of a filter first.

The last step in the training step is calculating normalization parameters for inference using the normalized spectra and the filters.
Furthermore, the normalized database vectors are calculated to which the inference vector is tested against to estimate the label.
The first step for classifying a spectrum is to record this spectrum using the filters which result from FBS.
Afterwards, the normalization parameters from the training step are applied to the recorded multispectral bands.
Finally, a spectral angle mapper (SAM) is used to classify the spectrum of interest.
A SAM calculates the spectral angle between the current vector to classify and each of the training vectors resulting from the last normalization step during training, and outputs the label with the minimal spectral angle.
The last normalization step in the training and in the inference is necessary since higher values, e.g. resulting from a higher filter bandwidth, would result in a higher influence on the spectral angle.

The same procedure is also done with the all other methods including two methods from the literature which do not rely on hyperspectral images, namely, maximizing orthogonality in characteristic reflectance vector space (MOCR) \cite{hardeberg_filter} and maximum linearity independence (MLI) \cite{li_filter_2018}.
For the full spectra, the FBS block is replaced by a block that outputs delta impulses as filters for each wavelength.
For the uniform filter selection, this block yields nine equidistant bandpass filters with a bandwidth of 50 nm, such that the whole wavelength area is covered.

The results of the classification procedure are summarized in Table \ref{tab:eval_classification}.
Our novel approach performs much better than the two methods from literature.
This result is expected since these methods follow a different goal.
MOCR tries to maximize the orthogonality in the space of characteristic reflectances.
This is nearly the inverse goal, since this method tries to find filters that capture movements most spectra have in common.
However, for our task it is necessary to filter out the regions that makes the spectra distinguishable.
MLI does not consider the underlying spectra at all but tries to find a filter configuration such that the filter matrix is well conditioned.
When not considering the underlying spectra, it is obvious that the most interesting regions for discrimination cannot be found.
The uniformly placed filters also have a worse performance than using the full spectra or FBS, which indicates that our smart filter selection procedure is actually improving the classification result.
FBS has even a slightly better performance in comparison to using the full spectra.
This is explainable by the averaging property of the filters which lowers the influence of noise on the classification.

\begin{table}[t]
	\centering
	\caption{The number of wrongly classified objects using filter selection methods from literature, uniformly placed filters, the filters chosen by FBS, and the full spectra.}
	\label{tab:eval_classification}
	\begin{tabular}{c|c|c|c||c}
		MOCR \cite{hardeberg_filter} & MLI \cite{li_filter_2018} & Uniform & FBS & Full spectra \\
		\hline
		459 & 388 & 350 & 318 & 320 \\
	\end{tabular}
\end{table}

\section{Conclusion}
\label{sec:conclusion}

This paper introduced a novel method to choose a subset of filters for a specific classification application using binary search.
It was shown that for all cases our novel approach reaches the optimum solution of a full search.
However, the full search solution is not always feasible since the number of possibilities is enormous already for a medium number of available filters.
In contrast, our fast binary search algorithm nearly takes a constant time for every desired number of filters.
The approximation procedure optimizes a mixed-integer program within a binary search over a desired metric.
The metric for the distance between two different filters can be changed arbitrarily within this framework.
The experiment also reveals a good performance for a real-world classification application using the set of filters chosen by our fast filter selection algorithm.
Future work will consider the influence of spectral imaging noise.

\bibliographystyle{ieeetr}
\bibliography{refs}
\vspace{12pt}

\end{document}